%% file: conference_101719.tex
\documentclass[conference]{IEEEtran}
\IEEEoverridecommandlockouts
\usepackage{cite}
\usepackage{amsmath,amssymb,amsfonts}
\usepackage{algorithmic}
\usepackage{graphicx}
\usepackage{textcomp}
\usepackage{xcolor}
\usepackage{acronym}
\usepackage{booktabs,tabularx,csquotes,url}
\usepackage{adjustbox}
\usepackage{multirow}
\usepackage{csquotes}
\usepackage{siunitx}
\usepackage{tikz,pgfplots}
\pgfplotsset{compat=1.18} 
\usepackage[colorlinks=false,pdfborder={0 0 0}]{hyperref}
\def\BibTeX{{\rm B\kern-.05em{\sc i\kern-.025em b}\kern-.08em
    T\kern-.1667em\lower.7ex\hbox{E}\kern-.125emX}}

\acrodef{ai}[AI]{artificial intelligence}
\acrodef{asr}[ASR]{automatic speech recognition}
\acrodef{comasa}[COMASA]{compact multiple
alignment for sequence averaging}
\acrodef{cnn}[CNN]{convolutional neural network}
\acrodef{ctc}[CTC]{connectionist temporal classification}
\acrodef{dtw}[DTW]{dynamic time warping}
\acrodef{dba}[DBA]{\acs{dtw} barycenter averaging}
\acrodef{hfcc}[HFCC]{human factor cepstral coefficient}
\acrodef{kws}[KWS]{keyword spotting}
\acrodef{mfcc}[MFCC]{Mel-frequency cepstral coefficient}
\acrodef{sed}[SED]{sound event detection}
\acrodef{stft}[STFT]{short-time Fourier transform}
\begin{document}

\title{Multi-Sample Dynamic Time Warping for Few-Shot Keyword Spotting}

\author{\IEEEauthorblockN{Kevin Wilkinghoff, Alessia Cornaggia-Urrigshardt}
\IEEEauthorblockA{\textit{Fraunhofer FKIE}\\
Fraunhoferstraße 20, 53343 Wachtberg, Germany\\
kevin.wilkinghoff@ieee.org, alessia.cornaggia-urrigshardt@fkie.fraunhofer.de}
}

\maketitle

\begin{abstract}
In multi-sample keyword spotting, each keyword class is represented by multiple spoken instances, called samples.
A na\"ive approach to detect keywords in a target sequence consists of querying all samples of all classes using sub-sequence dynamic time warping.
However, the resulting processing time increases linearly with respect to the number of samples belonging to each class.
Alternatively, only a single Fr\'echet mean can be queried for each class, resulting in reduced processing time but usually also in worse detection performance as the variability of the query samples is not captured sufficiently well. 
In this work, multi-sample dynamic time warping is proposed to compute class-specific cost-tensors that include the variability of all query samples.
To significantly reduce the computational complexity during inference, these cost tensors are converted to cost matrices before applying dynamic time warping.
In experimental evaluations for few-shot keyword spotting, it is shown that this method yields a very similar performance as using all individual query samples as templates while having a runtime that is only slightly slower than when using Fr\'echet means.
\end{abstract}


\section{Introduction}
\Ac{dtw} \cite{sakoe1978dynamic} and sub-sequence \ac{dtw} \cite{mueller2007information} generalize the Euclidean distance by allowing small temporal deviations when comparing sequential data and has many applications for classifying and clustering time series \cite{dau2019ucr}.
For audio data, a typical application of \ac{dtw} is \ac{sed}, i.e. detecting acoustic events in an audio signal with their on- and offsets in time while also correctly identifying the pre-defined classes they belong to.
Specific examples are bioacoustic event detection \cite{morfi2021few-shot,nolasco2022few-shot} and \ac{kws} \cite{lopez-espejo2020deep}.
Classically, hand-crafted speech features such as \acp{mfcc} or \acp{hfcc} extracted from isolated training samples are used as templates for sub-sequence \ac{dtw} \cite{von2010perceptual,kurth2010analysis}.
The same approach can also be used to obtain weakly labeled samples for training a discriminative model \cite{menon2018fast}.
More recently, neural networks have been applied to learn discriminative embeddings with a temporal resolution to be used as templates for \ac{dtw} yielding significantly better performance \cite{wilkinghoff2021twodimensional,wilkinghoff2024tacos}, even when only a very limited number of training samples is available (few-shot learning \cite{wang2021generalizing_few}).
Another modern approach is to approximate \ac{dtw} by soft-\ac{dtw} \cite{cuturi2017soft-dtw}, which is differentiable and thus can be used as a loss function for training neural networks, for applications such as multi-pitch estimation \cite{krause2023soft} in music information retrieval \cite{mueller2015fundamentals}.
\par
\Ac{dtw} has a computational complexity of $\mathcal{O}(N\cdot M)$ where $N\in\mathbb{N}$ and $M\in\mathbb{N}$ denote the lengths of the two sequences to be aligned.
This runtime can be improved by masking large parts of the cost matrices \cite{itakura1975minimum,sakoe1978dynamic}, pruning, i.e. not following, non-promising \ac{dtw} paths \cite{silva2016speeding} or parallelizing the computation of the accumulated cost matrices by sweeping diagonally over the entries \cite{zhu2018developing}.
Still, when applying \ac{dtw} to detect events of $C\in\mathbb{N}$ classes with $K\in\mathbb{N}$ query samples per class, the computational complexity increases to $\mathcal{O}(N\cdot M\cdot C\cdot K)$, which may be highly impractical or even infeasible.
To solve this computational issue, the Fr\'echet means can be computed from all samples of individual classes.
This is known to be an NP-complete problem \cite{buchin2022approximating} but heuristic approximation algorithms that work well in practice \cite{bruening2024number} exist.
Examples are \ac{dba} \cite{petitjean2011global} or \ac{comasa} \cite{petitjean2012summarizing}.
Although this reduces the complexity to $\mathcal{O}(N\cdot M\cdot C)$, using the Fr\'echet means as query samples also degrades the performance as the variability of the individual samples is not captured sufficiently well.
In \cite{petitjean2014dynamic,petitjean2016faster}, it has been proposed to cluster the individual samples and then compute Fr\'echet means for each cluster to essentially reduce the number of query samples to the number of clusters per class.
Still, multiple templates are needed for each class and the variability provided by the individual samples is not completely retained.
In this work, it is proposed to create class-specific cost tensors, which include the variability of all samples belonging to a class, and convert this tensor into a cost matrix before applying \ac{dtw}.
This procedure allows to efficiently compute the similarity between multiple query samples and a target sequence.
\par
The contributions of this work are the following.
First and foremost, multi-sample \ac{dtw}, a method that utilizes cost tensors to capture the variability provided by multiple query samples when computing class-specific \ac{dtw} matching scores, is presented.
In experiments conducted on \ac{kws}-DailyTalk, a dataset for few-shot open-set \ac{kws}, it is shown that multi-sample \ac{dtw} yields a similar performance as searching for all samples provided for a given class while still being computationally efficient.
In addition, a procedure to compute an altered Fr\'echet mean is proposed that leads to significantly better performance than the standard Fr\'echet mean.

\begin{figure*}
	\centering
	\begin{adjustbox}{max width=0.9\textwidth}
        \includegraphics{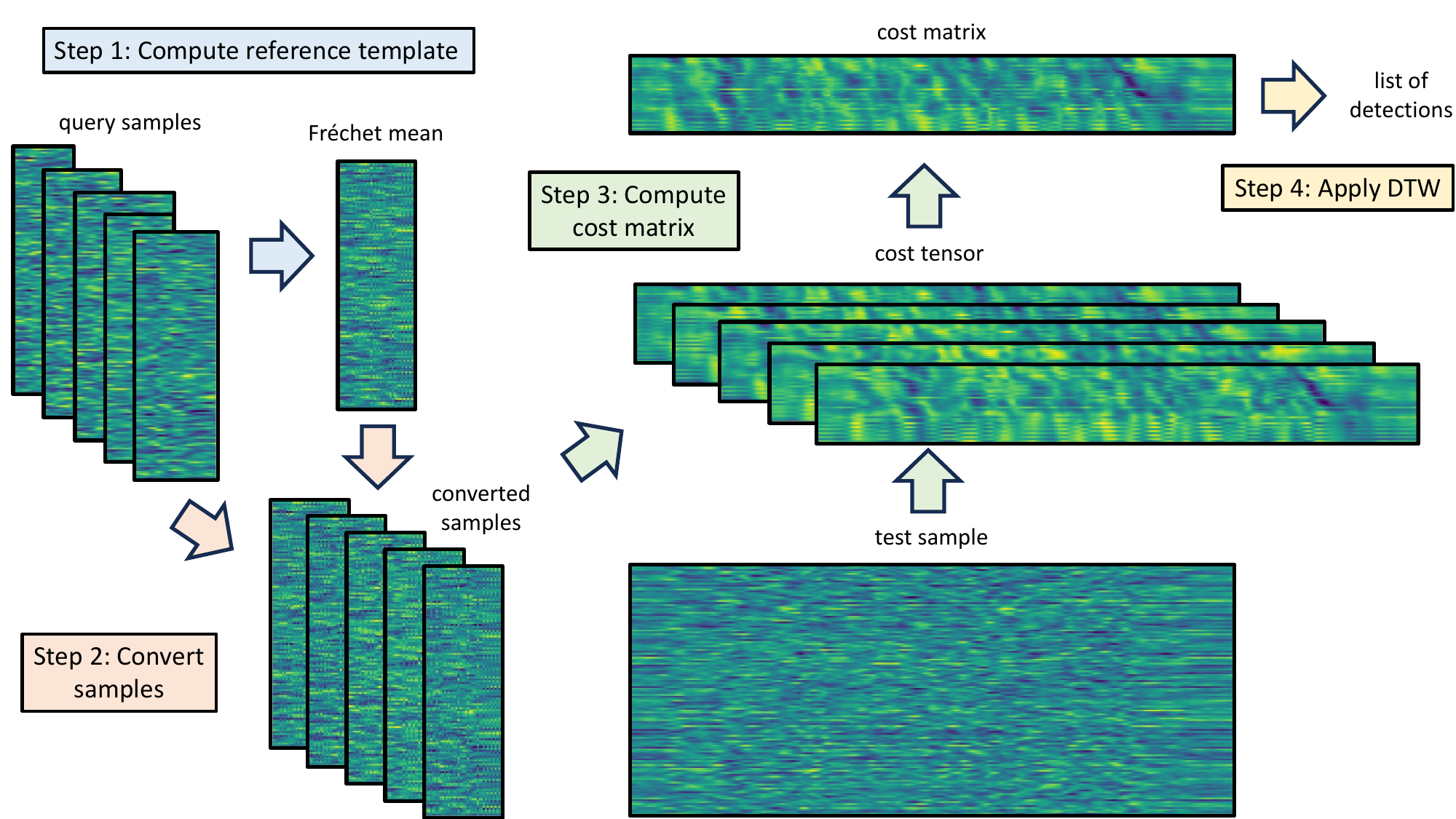}
	\end{adjustbox}
     \caption{Illustration of all steps of multi-sample \ac{dtw} for a single class using templates based on discriminative embeddings.}
	\label{fig:msdtw_illustration}
\end{figure*}
\section{Multi-sample \acs{dtw}}
\subsection{Definition}
\label{subsec:definition}
The idea of multi-sample \ac{dtw} is to allow the \ac{dtw} paths to switch between the provided query samples.
This is accomplished by combining the cost matrices belonging to the query samples of each class into class-specific cost tensors, which are then converted into cost matrices before applying \ac{dtw}.
More concretely, multi-sample \ac{dtw} consists of the following four steps, which are also depicted in \autoref{fig:msdtw_illustration}:
\par
\textbf{Step 1:} First, the Fr\'echet means are determined for each class.
These means serve as class-specific reference templates that are used to establish connections between the cost matrices of all query samples belonging to the same class.
For the experiments conducted in this work, the Fr\'echet means were initialized by applying the \ac{dba} algorithm \cite{petitjean2011global} based on \cite{schultz2018nonsmooth} as implemented in ts-learn \cite{tavenard2020tslearn} using a Sakoe-Chiba band \cite{sakoe1978dynamic} with a radius of $1$ and fixed sizes equal to the mean size of the corresponding query samples.
\par
\begin{figure}
	\centering
	\begin{adjustbox}{max width=\columnwidth}
        \includegraphics{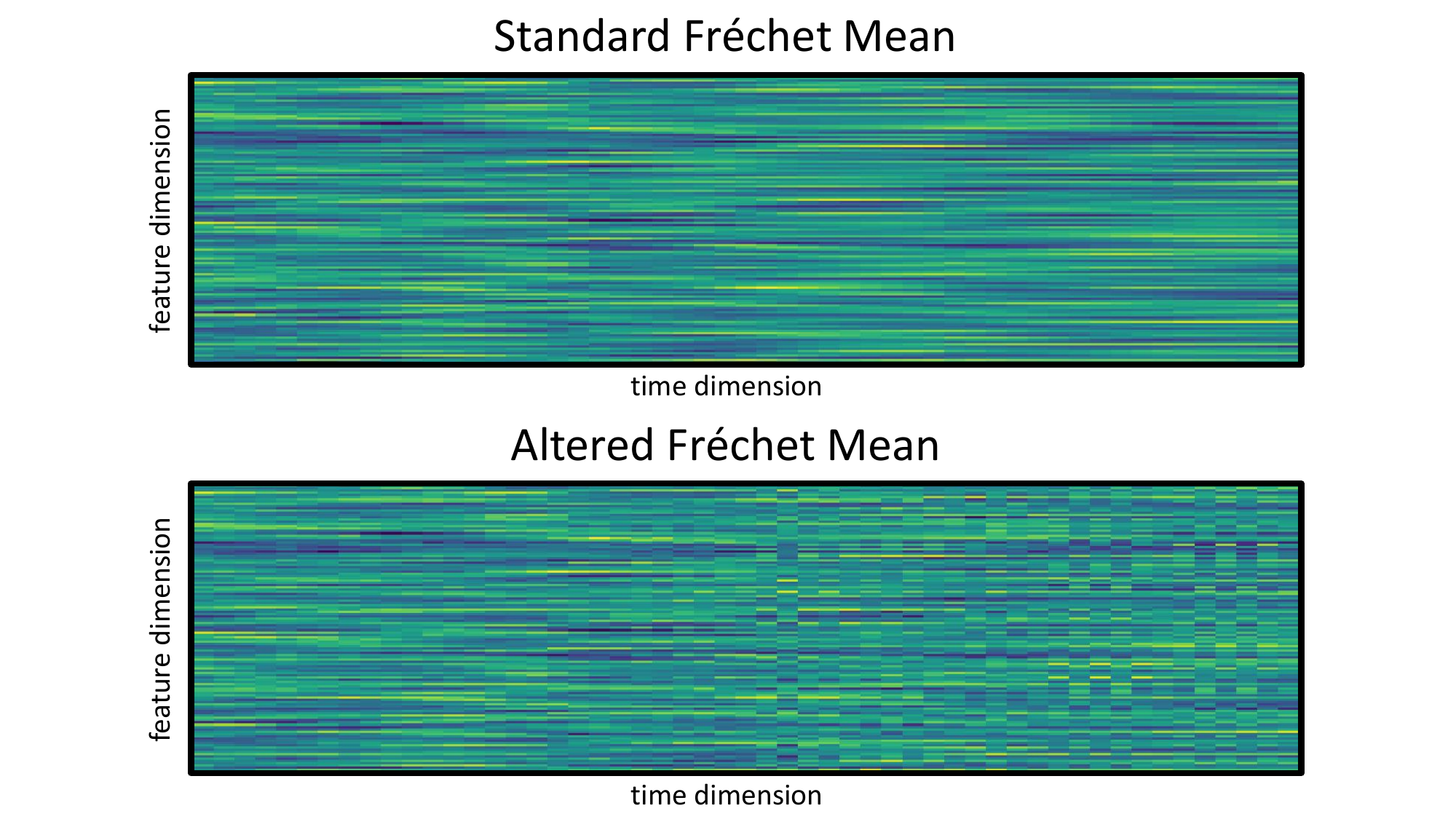}
	\end{adjustbox}
     \caption{Examples of a standard and an altered Fr\'echet mean based on discriminative embeddings. While both representations look very similar in the left half, the altered Fr\'echet mean alternates between the entries belonging to the standard mean and other entries in the right half. This allows to represent two alternative templates within a single template and thus captures more variability than the standard Fr\'echet mean.}
	\label{fig:means}
\end{figure}

The resulting standard Fr\'echet means are altered by applying the \ac{dba} algorithm a second time with sub-sequence \ac{dtw} and the non-standard steps $(1,1)$, $(1,2)$ and $(2,1)$.
The cosine distance is used as the local distance measure.
These steps allow ignoring every other entry when matching two sequences.
As a result, some variability of the query samples can be incorporated into the mean by alternating original entries with novel entries resulting from the second \ac{dba} algorithm.
This makes it possible to essentially capture two alternative templates within a single template.
An example for illustration purposes can be found in \autoref{fig:means}.
Note that in general the algorithm for computing the Fr\'echet mean can be chosen arbitrarily when using multi-sample \ac{dtw} and this particular choice was used as it significantly improved the performance (cf. \autoref{tab:performance}).
\par
\textbf{Step 2:} Second, all query samples are converted to have the same temporal dimension as the reference template of the class they belong to.
The converted sample is created from the reference template by replacing each entry individually matching each query sample with its corresponding reference template and replacing each entry of the reference template with the arithmetic mean of all entries that are matched to this position.
In other words, for each individual query sample a single iteration of the \ac{dba} algorithm is applied to modify the previously computed reference template.
For the experiments conducted in this work, sub-sequence \ac{dtw} with the cosine similarity and the non-standard steps $(1,1)$, $(1,2)$ and $(2,1)$ was used.
\par
\textbf{Step 3:} To obtain a single cost matrix describing the similarity between all converted query samples and a test sample, first a three-dimensional cost-tensor is computed by combining the cost matrices between the modified samples and the target sequence.
After that, the cost tensor is transformed into a standard cost matrix by taking the element-wise minimum over all cost matrices corresponding to slices of the cost tensor.
This allows finding \ac{dtw} paths of minimal cost that may switch between the cost matrices of different query samples. 
\par
\textbf{Step 4:} As a last step, standard (sub-sequence) \ac{dtw} can be applied to each class-specific cost matrix to obtain a similarity score for each class.
Hence, this step is the same as when utilizing Fr\'echet means as query samples.

\subsection{Runtime analysis}
One of the major goals of introducing multi-sample \ac{dtw} is to reduce the runtime during inference over computing \ac{dtw} similarity scores for all individual query samples.
Therefore, the computational complexity of multi-sample \ac{dtw} will now be investigated in more detail.
Computing the Fr\'echet means in steps 1 and converting the templates to have the same length as their corresponding mean in step 2 is only done once and thus can be considered as part of training
Step 4 and thus also its runtime is the same as when using class-specific Fr\'echet means and thus any runtime improvements for this step do not yield an advantage over the runtime needed when using the Fr\'echet means, which can be assumed to be optimal.
This means that the runtime is significantly faster than using the individual query samples because \ac{dtw} only needs to be applied to a single cost matrix for each class.
\par
Step 3 requires a more detailed inspection of its computational complexity:
Computing the cost matrices for each query sample has the same runtime as when using all individual query samples and thus is more costly than when only using the Fr\'echet means.
Furthermore, converting the cost tensor into a cost matrix by taking the element-wise minimum is an additional step that is relatively expensive but can be easily parallelized.
In the experiments conducted in this work, a sequential as well as a parallelized conversion of the cost tensor into a cost matrix will be compared.
\par
In conclusion, the runtime of multi-sample \ac{dtw} is much faster than when using all individual query samples because only one cost matrix is generated for each class.
Thus, after extracting the cost matrix the computational complexity is reduced from $\mathcal{O}(N\cdot M\cdot C\cdot K)$ to $\mathcal{O}(N\cdot M\cdot C)$.
Yet, the runtime is slower than only using a single Fr\'echet mean for each class because computing this cost matrix requires to compute a cost tensor, which has a computational complexity of $\mathcal{O}(N\cdot M\cdot C\cdot K)$.
In total, the computational complexity is still in $\mathcal{O}(N\cdot M\cdot C\cdot K)$ but with a much smaller constant than when using the individual query samples.

\section{Experimental evaluation}

\subsection{Dataset}
For all experiments conducted in this work, the few-shot open-set \ac{kws} dataset \ac{kws}-DailyTalk \cite{wilkinghoff2024tacos} based on the speech dataset DailyTalk \cite{lee2022dailytalk} was used.
\Ac{kws}-DailyTalk consists of a training split as well as a validation and a test split.
The training split contains five isolated samples for each of the fifteen keywords \enquote{afternoon}, \enquote{airport}, \enquote{cash}, \enquote{credit card}, \enquote{deposit}, \enquote{dollar}, \enquote{evening}, \enquote{expensive}, \enquote{house}, \enquote{information}, \enquote{money}, \enquote{morning}, \enquote{night}, \enquote{visa} and \enquote{yuan}.
The validation and the test split consist of $156$ and $157$ sentences, respectively, that contain either none or several of these keywords such that each keyword appears about twelve times in each of the splits.
Each training sample is taken from other conversations than the sentences contained in the validation or test split.
In total, the duration of the training split is \SI{39}{\second} and both other splits each have a duration of approximately \SI{10}{\minute}.
To evaluate the performance, the micro-averaged event-based F-Score as implemented in the \ac{sed} evaluation package \cite{mesaros2016metrics} is used.

\subsection{Baseline systems}
In order to compare the performance and runtime obtained with the proposed multi-sample \ac{dtw} approach to using all individual samples or the Fr\'echet means as query samples, the following two baseline systems based on different input features were used:
\par
\textbf{\acp{hfcc}:} 
As a first \ac{dtw} feature, \acp{hfcc}, which are known to outperform other hand-crafted speech features such as \acp{mfcc} for query-by-example \ac{kws} \cite{von2010perceptual,kurth2010analysis}, were used.
In contrast to embeddings extract with neural networks, these features have the advantage that they do not require any training and thus can be used as an ad-hoc \ac{kws} approach.
The \acp{hfcc} were calculated based on spectrograms with a window size of $\SI{25}{\milli\second}$ and a step size of $\SI{10}{\milli\second}$. 
\par
\textbf{Discriminative Embeddings:} 
As a second \ac{dtw} feature, embeddings with a temporal dimension obtained by using the embedding model based on the TACos loss function as proposed in \cite{wilkinghoff2024tacos} were used.
These embeddings are obtained by using spectrograms of short windows with a length of \SI{250}{\milli\second} and an overlap of \SI{50}{\milli\second} as input to a convolutional neural network and have the same time dimension as the input spectrogram.
The network is trained to not only predict the keyword a given segment belongs to but also to predict the position of the segment inside the keyword recording by minimizing an angular margin loss.
After training the network, audio recordings can be converted into embeddings by dividing them into segments, computing the embeddings and taking the mean of all frames at positions that overlap in time.
This results in learning discriminative embeddings that are well-suited for predicting the correct keyword and change over time to be suitable to be used as features for \ac{dtw}.
Furthermore, temporally reversed segments are used as additional challenging negative examples for each keyword to reduce the number of false alarms.
More details about the loss function as well as the \ac{kws} system can be found in \cite{wilkinghoff2024tacos}.
\begin{table*}
	\centering
	\caption{Event-based, micro-averaged F-score, precision and recall as well as runtimes obtained with different \ac{dtw} approaches on the validation and test split of \ac{kws}-DailyTalk using all five shots. Highest F-scores for each \ac{dtw} feature and lowest runtimes are highlighted with bold letters. The runtime needed to apply the \ac{dtw} approaches after extracting all features is measured on an Intel i7-8700 CPU @\SI{3.20}{\giga\hertz}.}
\begin{adjustbox}{max width=\textwidth}
	\begin{tabular}{lllll|lll|r}
		\toprule
		\pmb{\ac{dtw} feature}&\pmb{\ac{dtw} approach}&\multicolumn{6}{c}{\pmb{obtained performance}}&\pmb{runtime}\\
        &&\multicolumn{3}{c}{\pmb{validation set}}&\multicolumn{3}{c}{\pmb{test set}}&\\
        &&\pmb{F-score}&\pmb{precision}&\pmb{recall}&\pmb{F-score}&\pmb{precision}&\pmb{recall}&\\
		\midrule
        \acp{hfcc}&\ac{dtw} with individual samples&\pmb{60.18\%}&64.56\%&56.35\%&\pmb{58.82\%}&62.89\%&55.25\%&\SI{29.5}{\second}\\
        \acp{hfcc}&\ac{dtw} with standard Fr\'echet means&45.00\%&51.80\%&39.78\%&42.44\%&50.77\%&36.46\%&\pmb{\SI{6.2}{\second}}\\
        \acp{hfcc}&\ac{dtw} with altered Fr\'echet means&51.10\%&59.56\%&44.75\%&49.54\%&55.48\%&44.75\%&\pmb{\SI{6.2}{\second}}\\
        \acp{hfcc}&multi-sample \ac{dtw}&57.93\%&64.63\%&52.49\%&58.08\%&63.40\%&53.59\%&\SI{20.9}{\second}\\
        \acp{hfcc}&multi-sample \ac{dtw} (parallelized)&57.93\%&64.63\%&52.49\%&58.08\%&63.40\%&53.59\%&\SI{12.4}{\second}\\
		\midrule
        embeddings&\ac{dtw} with individual samples&69.18\%&80.29\%&60.77\%&\pmb{69.48\%}&84.25\%&59.12\%&\SI{18.9}{\second}\\
        embeddings&\ac{dtw} with standard Fr\'echet means&66.24\%&79.23\%&56.91\%&59.26\%&75.86\%&48.62\%&\pmb{\SI{4.2}{\second}}\\
        embeddings&\ac{dtw} with altered Fr\'echet means&68.94\%&78.72\%&61.33\%&67.73\%&80.30\%&58.56\%&\pmb{\SI{4.2}{\second}}\\
        embeddings&multi-sample \ac{dtw}&\pmb{71.07\%}&82.48\%&62.43\%&69.43\%&81.95\%&60.22\%&\SI{10.6}{\second}\\
        embeddings&multi-sample \ac{dtw} (parallelized)&\pmb{71.07\%}&82.48\%&62.43\%&69.43\%&81.95\%&60.22\%&\SI{6.5}{\second}\\
		\bottomrule
	\end{tabular}
\end{adjustbox}
\label{tab:performance}
\end{table*}
\par
For both systems, sub-sequence \ac{dtw} with the cosine similarity and the non-standard steps $(1,1)$, $(1,2)$ and $(2,1)$ was used.
Furthermore, the detection results were post-processed by shortening all overlapping events such that only the event with the highest \ac{dtw} similarity score is kept and removing all detections that are shorter than $50\%$ of the length of the corresponding reference template.
All hyperparameters such as the decision thresholds of the \ac{kws} systems were optimized with respect to the performance on the validation split.
The Fr\'echet means are calculated as described in step 1 of Section \ref{subsec:definition}.

\subsection{Experimental results}
\label{sec:exp_results}
The experimental results obtained on \ac{kws}-DailyTalk are presented in \autoref{tab:performance}. 
First, it can be seen that the embeddings outperform the \acp{hfcc} in all experiments, as also shown in \cite{wilkinghoff2024tacos}.
Furthermore, using the Fr\'echet means as query samples significantly degrades the performance obtained when using all individual samples and the altered Fr\'echet means perform much better than the standard Fr\'echet means.
This is especially true for the case of the \acp{hfcc} that are not trained discriminatively to have high intra-class similarity and low inter-class similarity.
Therefore, the means indeed do not capture the variability of the individual samples caused by different speaker characteristics and the context sufficiently well.
\par
Another major observation is that multi-sample \ac{dtw} leads to a significantly better performance than using the Fr\'echet means as reference templates, regardless of the input feature being used.
Moreover, with multi-sample \ac{dtw} the performance is close to the one obtained using individual samples, sometimes even slightly better, while being much faster than the latter one.
This is due to the fact that with the proposed method new artificial samples are generated by enabling the use of any combination of templates.
\par
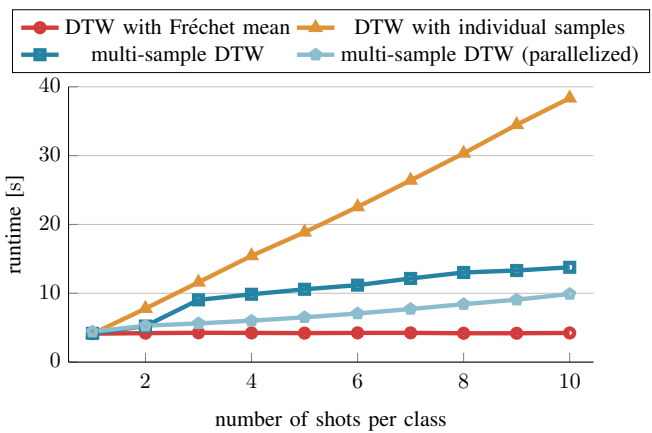
\begin{figure}
    \centering
    \begin{adjustbox}{max width=0.99\columnwidth}
          \input{images/comparison.tikz}
    \end{adjustbox}
    \caption{Runtime with respect to the number of shots for each class when applying different \acs{dtw} approaches to the embeddings. All runtimes of the \ac{dtw} approaches are measured after extracting all features using an Intel i7-8700 CPU @\SI{3.20}{\giga\hertz}.}
    \label{fig:comparison}
\end{figure}

A detailed comparison of the runtimes obtained with the different approaches can be found in \autoref{fig:comparison}.
As expected, the lowest runtime is achieved when using the Fr\'echet means.
The slowest solution is to use individual samples as the runtime increases linearly with the number of shots per class and thus this approach may be infeasible.
The runtime of multi-sample \ac{dtw} is much better than using all individual query samples because applying \ac{dtw} to the cost matrices belonging to the relatively long test samples, which is the computationally most expensive step, only needs to be applied once per class.
Furthermore, converting the cost tensors into cost matrices by finding the minimum for all entries can be parallelized, which leads to an even better runtime as shown by the light blue line in \autoref{fig:comparison}.
As a result, a runtime very close to the one resulting from using the Fr\'echet means as query samples, which can be assumed to be optimal, is obtained.
It is worth noting that a higher recall means that more \ac{dtw} paths need to be investigated and this also increases the runtime.
Last but not least, the runtime of all approaches can be further reduced by also parallelizing the computation of the cost matrices or using more powerful hardware such as graphical processing units.
As this has the same impact for all approaches, it is not necessary when comparing their runtime.

\section{Conclusion and Future Work}
In this work, multi-sample \ac{dtw}, a method for efficiently computing the \ac{dtw} matching score of multiple samples belonging to the same class, was proposed.
This method consists of the following steps:
First, a reference template is determined for each class by computing the Fr\'echet means of the corresponding query samples.
Second, each query sample is matched with its reference template and the resulting assigment is used to convert the sample to have the same temporal dimension as the reference template.
All the modified samples of one class are then used to compute a three-dimensional cost tensor, which is converted to a standard cost matrix by taking the minimum over the sample dimension.
Then, \ac{dtw} can be applied to these class-specific cost matrices.
\par
The proposed multi-sample \ac{dtw} approach has been evaluated for the task of few-shot \ac{kws} on \ac{kws}-DailyTalk.
The experimental results show that using multi-sample \ac{dtw} yields the same performance as using every individual sample, which is much better as the performance obtained when using Fr\'echet means as query samples.
Furthermore, in contrast to using all individual query samples, the runtime only slightly increases with the number of samples and thus is still reasonably close to the optimal runtime.
The reason is that \ac{dtw} has to be applied only once per class and the step of converting the cost tensors into cost matrices has been parallelized.
As another contribution, a two-step procedure for significantly improving the performance of the Fr\'echet mean was proposed.
\par
For future work, it is planned to evaluate the proposed multi-sample \ac{dtw} approach on other \ac{kws} datasets as for example Speech Commands \cite{warden2018speech} and for other applications of \ac{dtw}.

\section{Acknowledgments}
The authors would like to thank Paul M. Baggenstoss, Frank Kurth and Sebastian Urrigshardt for their valuable feedback.

\bibliographystyle{IEEEtran}
\bibliography{mybib}

\end{document}

%% file: images/comparison.tikz
\begin{tikzpicture}
\begin{axis}[
	axis y line*=left,
    axis x line*=bottom,
    ymin=0.0,
    ymax=40,
    xmin=1,
    xmax=10,
    enlarge x limits=0.05,
    legend style={at={(0.5,1.05)},anchor=south,legend columns=2},
    ylabel=runtime \lbrack \si{\second}\rbrack,
    xlabel=number of shots per class,
    height=6cm,
    width=10cm,
    xticklabel style={align=center},
    yticklabel style={align=center},
    typeset ticklabels with strut,
    xlabel near ticks,
    ylabel near ticks,
    nodes near coords style={/pgf/number format/.cd,fixed zerofill,precision=2},
    ymajorgrids
]
\addplot[teal!20!red!85,mark=o,line width=2pt] coordinates {(1,4.131)(2,4.209)(3,4.263)(4,4.243)(5,4.210)(6,4.247)(7,4.251)(8,4.192)(9,4.194)(10,4.237)}; 
\addplot[teal!15!orange!85,mark=triangle,line width=2pt] coordinates {(1,4.104)(2,7.779)(3,11.598)(4,15.442)(5,18.867)(6,22.548)(7,26.401)(8,30.334)(9,34.487)(10,38.345)}; 
\addplot[teal!85!blue!85,mark=square,line width=2pt] coordinates {(1,4.180)(2,5.238)(3,9.042)(4,9.855)(5,10.580)(6,11.183)(7,12.153)(8,13.023)(9,13.311)(10,13.776)}; 
\addplot[teal!85!blue!45,mark=pentagon,line width=2pt] coordinates {(1,4.418)(2,5.289)(3,5.629)(4,6.009)(5,6.513)(6,7.063)(7,7.716)(8,8.418)(9,9.068)(10,9.899)}; 
\legend{\acs{dtw} with Fr\'echet mean, \acs{dtw} with individual samples, multi-sample \acs{dtw}, multi-sample \acs{dtw} (parallelized)}
\end{axis}
\end{tikzpicture}